\documentclass[slac_one]{revtex4}
\usepackage{graphicx}
\usepackage{fancyhdr}
\begin{document}

\title{Detection of new states using forward proton tagging at the LHC} 
%
\author{P. J. Bussey}
\affiliation{Department of Physics and Astronomy, Faculty of Physical Sciences\\
University of Glasgow, Glasgow G12 8QQ, U.K.\\ 
{Royal Society of Edinburgh / Scottish Executive Support Research Fellow}
\\ ~\\For the FP 420 Collaboration.
}
\begin{abstract}
~\\[-7mm] This talk summarises the ongoing proposals to upgrade the
ATLAS and CMS detectors by the installation of forward silicon
detector systems close to the beam line at distances of approximately
220 m and 420 m from the respective Interaction Points.  The physics
motivation is outlined, with emphasis on detection of Higgs and
Supersymmetric states, and some of the aspects of the apparatus and
its performance are briefly described.
\end{abstract}
\maketitle

\thispagestyle{fancy}

\section{Introduction and basic proposal for forward detectors.}
An important part of the programme of physics at HERA and the Tevatron
has been the measurement of diffractive processes, in which the proton
exchanges a colourless object, commonly referred to as the pomeron. Of
particular interest here are the production of exclusive final states,
such as vector mesons, and hard processes in which the partonic
components of the structure of the pomeron engage in the interaction.
The hard processes were induced at HERA by photons of varying
virtuality, ranging from quasi-real photons to highly virtual photons
giving deep inelastic scattering off the partons associated with the
pomeron.  At LHC, much higher energies are becoming available,
enabling the diffractive programme to be extended into areas where new
physics can be studied.  To do this it is proposed to install 
detector systems close to the beam line at
suitable locations downstream of the interaction points.  A summary is
presented here of some of the new processes that should be open to
investigation, and we finally return to outline the capabilities of
the physical apparatus in more detail.

The LHC beamline has separate incoming and outgoing beams, and at
distances greater than 260 m, the beam is steered by the main
bending magnets.  At two regions, namely around 220 m and 420 m from
the interaction point, there are intervals in the beamline 
unoccupied by magnets, each of which provides approximately ten
metres of clear space within which physics detectors can be stationed.
Sets of silicon detectors will be installed in these regions,
allowing them to approach as closely as possible to the outgoing beam.
These detectors will detect diffractively scattered outgoing protons.

One or both protons in a $pp$ collision may be scattered
diffractively.  The fractional energy loss $\xi$
suffered by the proton is typically small, as is the angle of
scatter. These protons will continue to travel along the beam line,
but eventually they will no longer be contained by the beam optics
and will be bent either into a collimator or out of the beam line
altogether.  It is found that protons that have lost a few tens of GeV
in the initial collision emerge out of the beam typically in the 420 m
regions, and those that have lost a few hundreds of GeV emerge in the
220 m regions.  By installing detector systems in these regions, we
can identify the double diffractive production of exclusive centrally
produced states whose mass is above a minimum value of the order of
100 GeV/c$^2$, provided that the state gives a suitable signature in the
central detector to allow its identification. 
A measurement of the energies of the outgoing protons gives a good
determination of the mass of the centrally produced object.

\section{Physics studies}
Standard Model Higgs production at the LHC has been
 calculated by a
number of authors.  The detected cross section of
 course depends on
the ability to trigger the process in the apparatus.
 Unfortunately
the present electronics in ATLAS and CMS do not
 allow a first-level
trigger to be based on a proton detection at 420
 m, since the signal
arrives too late.  This forces the detection of a
 120 GeV/c$^2$
central state to be based on central detector triggers,
 which are not
highly efficient for a SM Higgs at this mass.  In our
 favour is that
the background of quark-antiquark jets is suppressed by
 the $J_z=0$
selection rule \cite{KMR1}.  An exclusive
 double-diffractively
produced state is constrained to have $J^{PC} = 0^{++}$,
 so that if a
Higgs or other particle is seen at all in this process,
 we have a
good determination of its quantum numbers which may be hard
 to
determine unambiguously by central detector measurements alone.
 
\begin{figure}[b]
        \centering
        \includegraphics[width=0.47\textwidth]{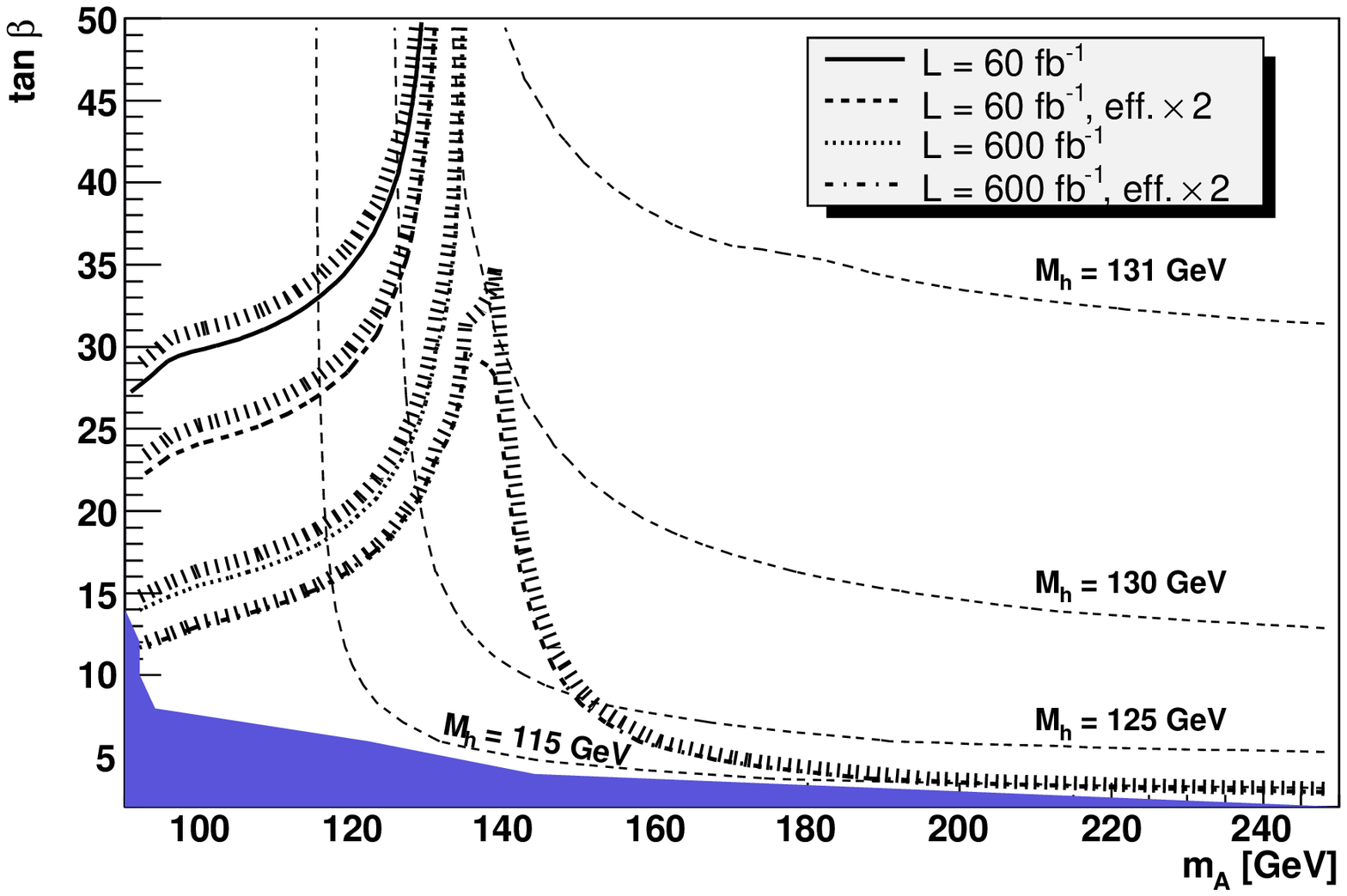}
        \hspace*{0.01\textwidth}
        \includegraphics[width=0.47\textwidth]{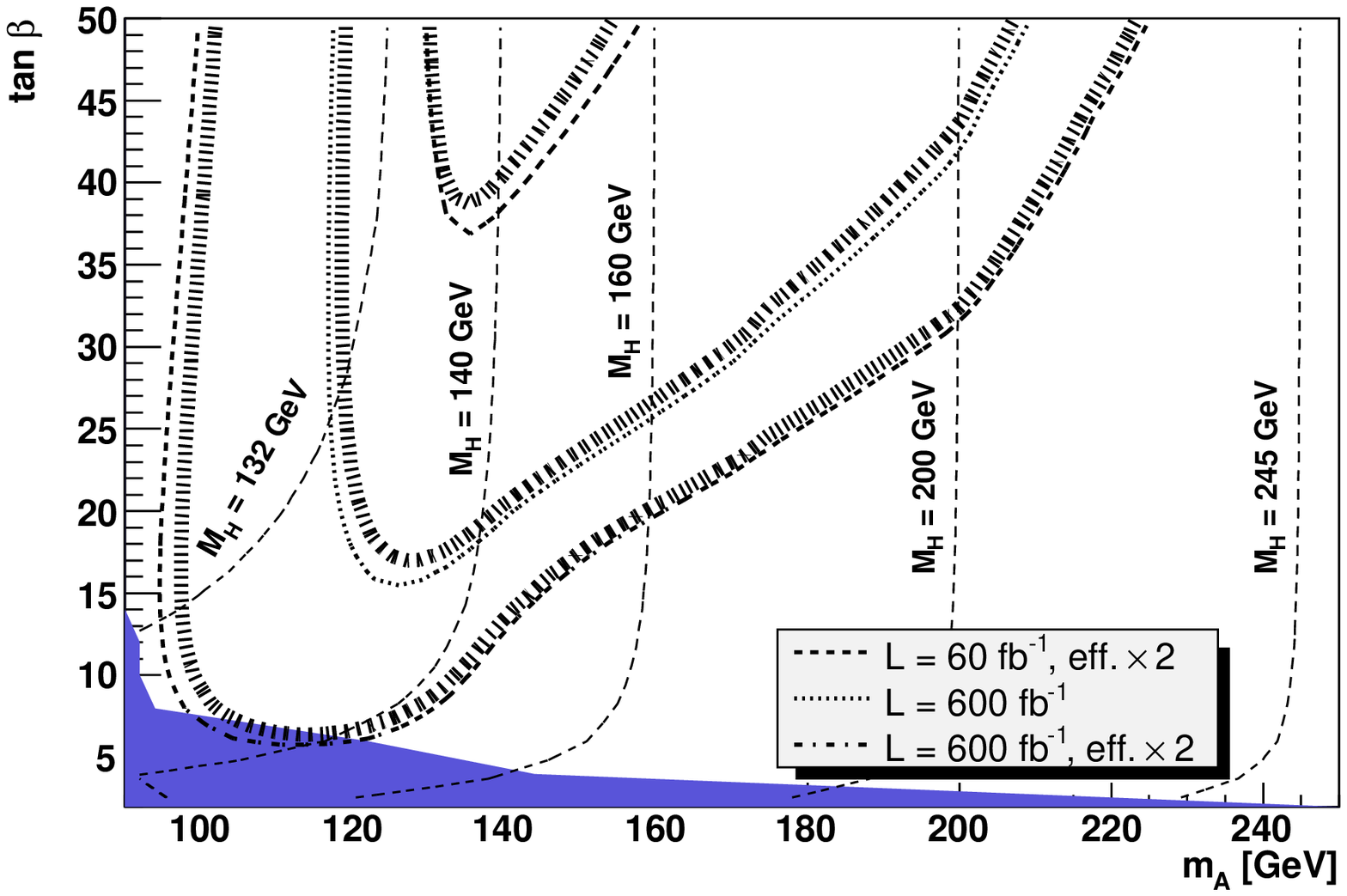}
        \caption{Contours for 3-$\sigma$ evidence (left) and 5-$\sigma$ discovery (right)
for the $h$ and $H$ SUSY Higgs (see text).
        }
        \label{fig:fwd_4}
\end{figure}

With the present set-up the prospects for SM Higgs detection in double
diffraction
 at the LHC seem rather marginal.  However there are
additional
 opportunities if the Higgs is found within a
supersymmetric framework.
 There are two particularly important
parameters of the SUSY scenario,
 denoted as $m_A$ and $\tan\beta$, in
whose parameter space a number of
 the features of the theory can be
illustrated.  In certain regions, enhancements to the SM Higgs cross
section might
 be obtained for the lighter of the two neutral SUSY
Higgs particles,
 denoted as $h$. 
On this basis, the
 quantity of LHC luminosity needed for 3-$\sigma$
evidence and
 5-$\sigma$ discovery of neutral SUSY Higgs in the
exclusive
 double-diffractive mode can be estimated, as illustrated
for the
 heavier SUSY Higgs $H$ in Fig.~\ref{fig:fwd_4}. Contour
plots of this kind have been presented by Heinemeyer et al.~\cite{Hein1} for the $h$
and $H$ in a variety of related situations. This gives improved hope
of being able to make Higgs studies with forward detectors at the
LHC,
 although there is no advance guarantee that the values of the
SUSY
 parameters will be favourable, and the integrated luminosity
needed
 might be substantial.
 
More cleverly thought-out triggers and cuts may improve the situation.
Pilkington et al.~\cite{Pilko} have reconstructed the mass of the
central object as reconstructed using modelled measurements of
the forward proton trajectories at 420 m, with estimated backgrounds
from other processes included.  During the first years of running, a
measurement using 60 fb$^{-1}$ seems a reasonable target. Higher
luminosities will clearly assist, but will generate combinatorial
backgrounds from overlapping events. If these can be removed, as is
envisaged, using precise timing measurements to isolate the event of
interest, a signal might be seen giving a 5-$\sigma$ discovery with
100 fb$^{-1}$ of running.

TeV-energy protons are surprisingly efficient at radiating high energy
photons.  Single photoproduction off the second proton and
photon-photon processes are both of interest at LHC.  Kinematically,
photoproduction resembles diffractive scattering but with the tendency
to smaller transverse momentum transfers to the proton.  Since
diffraction produces mainly gluon jets and photoproduction produces
quark jets, there is little interference between the processes.
Single photoproduction will be of interest at the LHC in the
production of electroweak particles. There are possibilities for the
associated production of Higgs bosons and for the production of
anomalous single top via FCNC.  These processes are tagged by a single
forward proton but must be triggered and identified in the central
detectors, and there will be potential difficulties at high
luminosities since the use of timing to associate the forward protons
with a central vertex requires two such forward protons.

The $\gamma\gamma$ process is capable of inducing the production of
any type of charged particle-antiparticle pair. Of particular interest
is the possible production of charged SUSY particles, such as
charginos and sleptons, whose signatures in the central detector will
be high transverse energy leptons and missing energy carried by
neutrinos or the lightest SUSY particle (LSP) if it is
neutral. $W^+W^-$ production is likely to be a very prolific 
background. 

\begin{figure}[b!]
        \centering
        \includegraphics[width=0.4\textwidth]{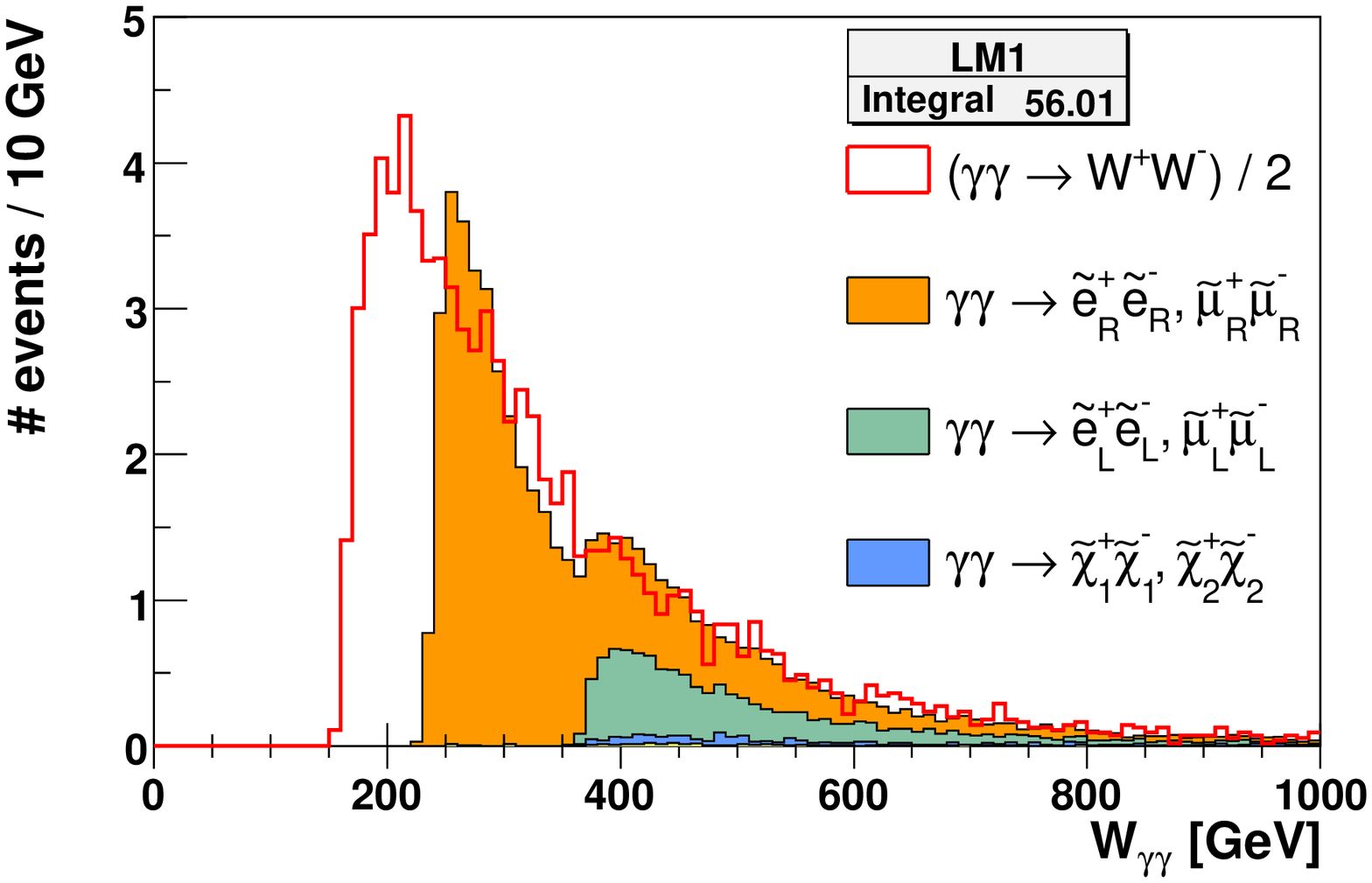}
        \hspace*{0.05\textwidth}
        \includegraphics[width=0.4\textwidth]{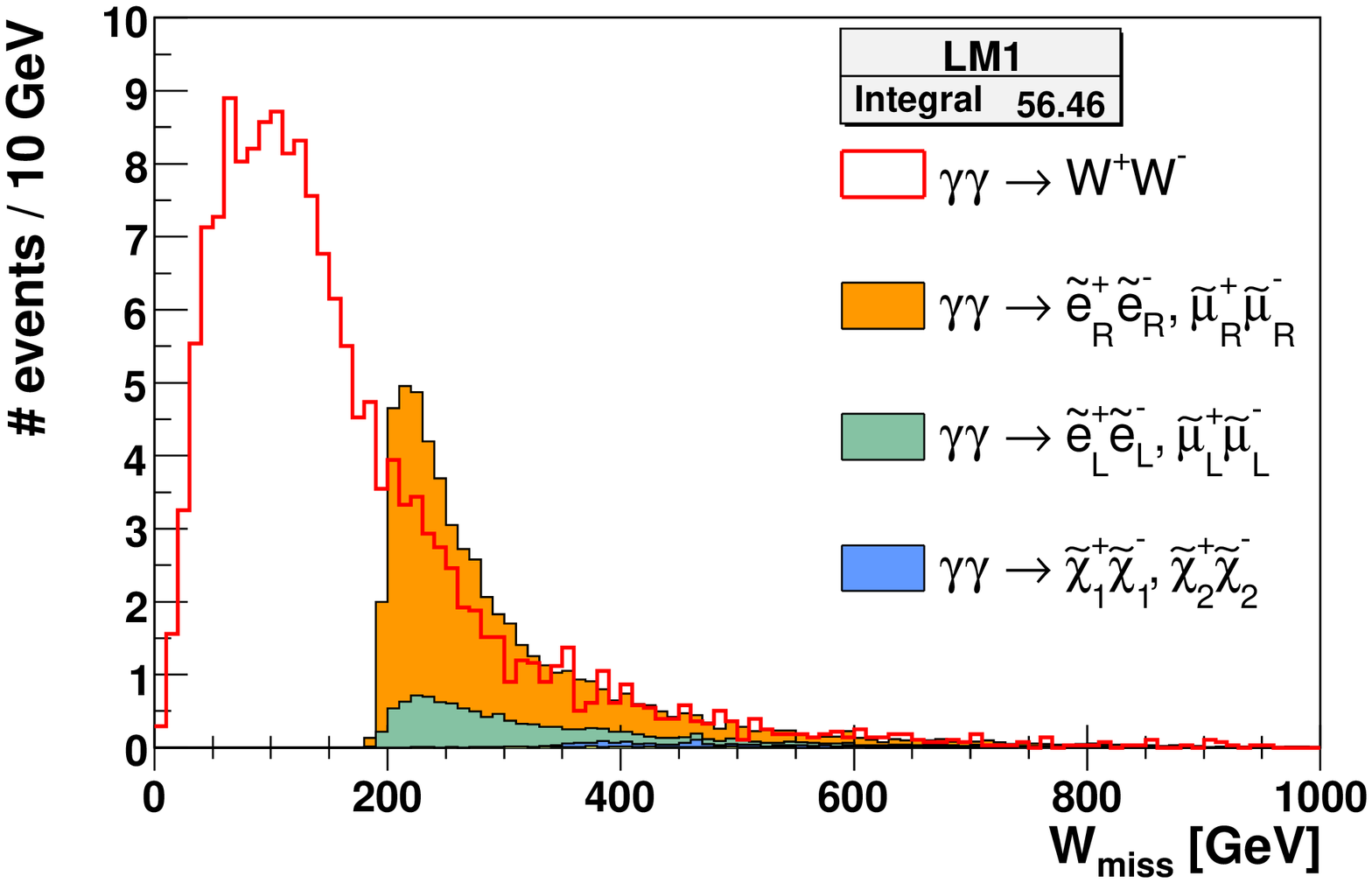}\\[-33.5ex]
        (a)\hspace*{0.45\textwidth}(b)\hspace*{0.24\textwidth}~\\[30ex]
        \caption{Examples of the analysis of the double photoproduction of SUSY
particles, as a function of the parameters $W_{\gamma\gamma}$ and $W_{miss}$,
to illustrate a possible way to isolate a clean SUSY signal (N. Schul)
        }
        \label{fig:fwd_7}
\end{figure}

There are many SUSY mass scenarios.  Some possibilities have been
studied here in terms of the so-called LM1 scenario, which involves a
light LSP and light sleptons and charginos.  This type of scenario
would give the most favourable set of cross sections.  The most
natural variable to plot in order to separate SUSY signals from $WW$
background would be the $W_{\gamma\gamma}$ value reconstructed from
the forward protons (Fig.~\ref{fig:fwd_7}a) However the background is
much more tractable when the variable $W_{miss}=\sqrt{E^2_{miss} -
P^2_{miss}}$ is plotted (Fig.~\ref{fig:fwd_7}b), where the missing
energy and momentum are easy calculated from the forward protons and
the kinematics of the observed final state particles. Combinations of
$W_{\gamma\gamma}$ and $W_{miss}$ give even more power and can
generate a distribution that might give a 5-$\sigma$ discovery with
only 25 fb$^{-1}$ of integrated luminosity.

An extended range of SUSY processes may be accessible.  One 
that has been studied is the detection of pairs of long-lived gluinos,
for which the forward detectors at 220m and 420 m give access to the
wide range of masses that such particle pairs may have~\cite{Cough}. An
intriguing example of completely new physics has been proposed by
A. White in which a new SU(5) gauge theory obviates the need for a
Higgs particle and gives remarkable experimental signatures for which
pomeron physics may be an essential diagnostic tool \cite{White}.

Space is too limited here to mention more than briefly other items
in the range of physics processes that will be made observable by the
use of forward tagging systems.  The work initiated at HERA on hard
pomeron scattering and structure can be continued at LHC by means of
photon-pomeron and pomeron-pomeron processes.  There will be extended
opportunities for further studies of the nature of the pomeron. In the
early stages, at low LHC luminosities, the study of rapidity-gap
survival will be interesting and important, generalised gluon
distributions can be studied, and a variety of QCD effects can be
investigated; a recent review by Khoze, Martin and Ryskin gives more
details~\cite{KMR}.

\section{The proposed apparatus}
The traditional idea of Roman Pots has been extended so that we plan
to have an entire section of movable beam pipe, the so-called
``Hamburg Pipe'', within which sets of silicon detectors will be
mounted. The cryostat connection between the portions of beamline
either side of the 420 m installations must be replaced. Two optimise
performance, two sets of detectors are installed in each pipe,
separated by approximately 10 m,
so that the position and angle of an emerging
proton trajectory can be measured. In the horizontal plane, precisions of
approximately 10 $\mu$m in position and 1 $\mu$rad in angle should be
obtainable. Reduced precision in the less critical vertical plane
will be accepted.  The silicon detectors are of a recent ``edgeless''
technology to allow the sensitive area to be moved as close as
possible to the main outgoing proton beam.

\begin{figure}[t!]
        \centering
        \includegraphics[width=0.4\textwidth]{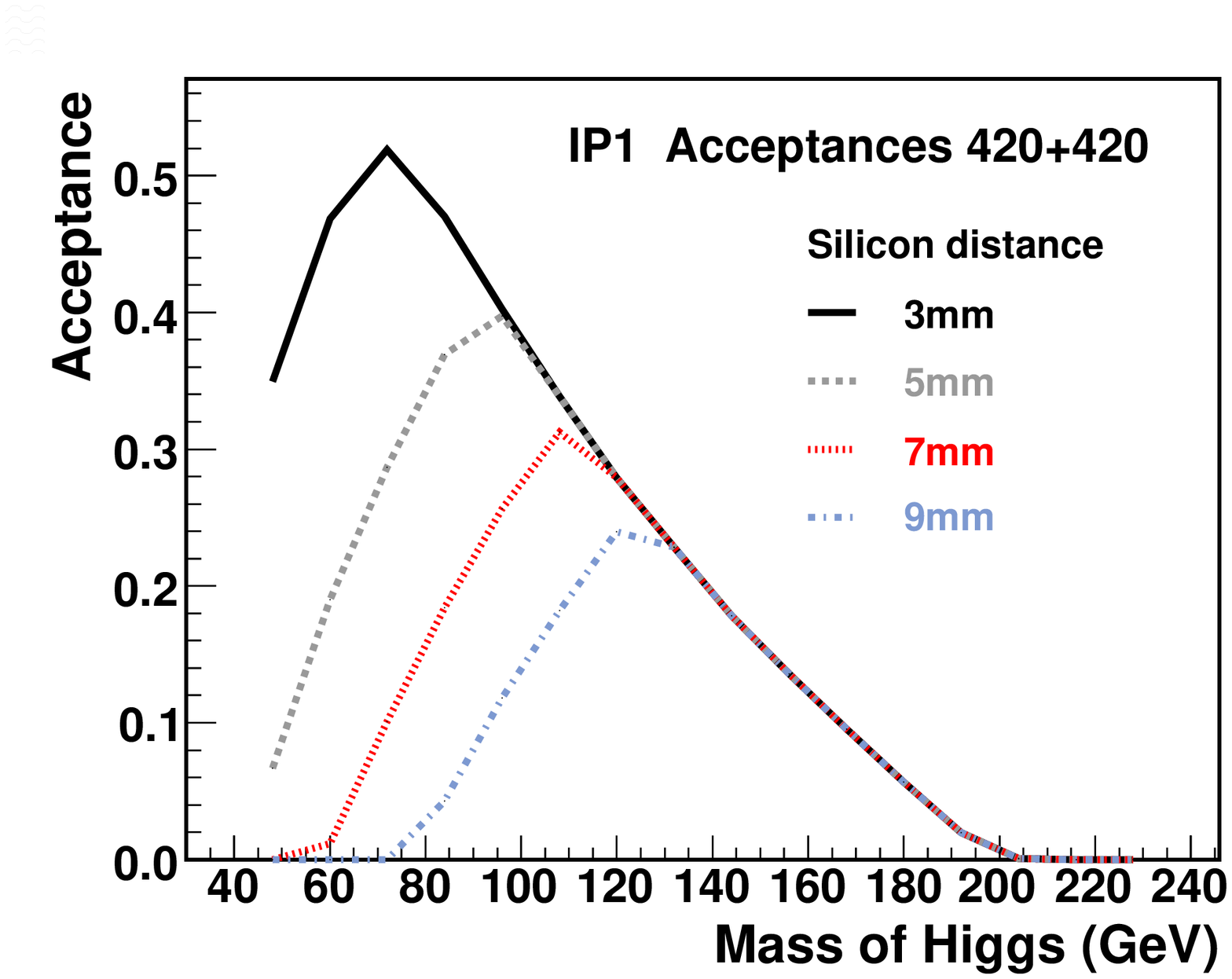}
        \hspace*{0.05\textwidth}
        \includegraphics[width=0.4\textwidth]{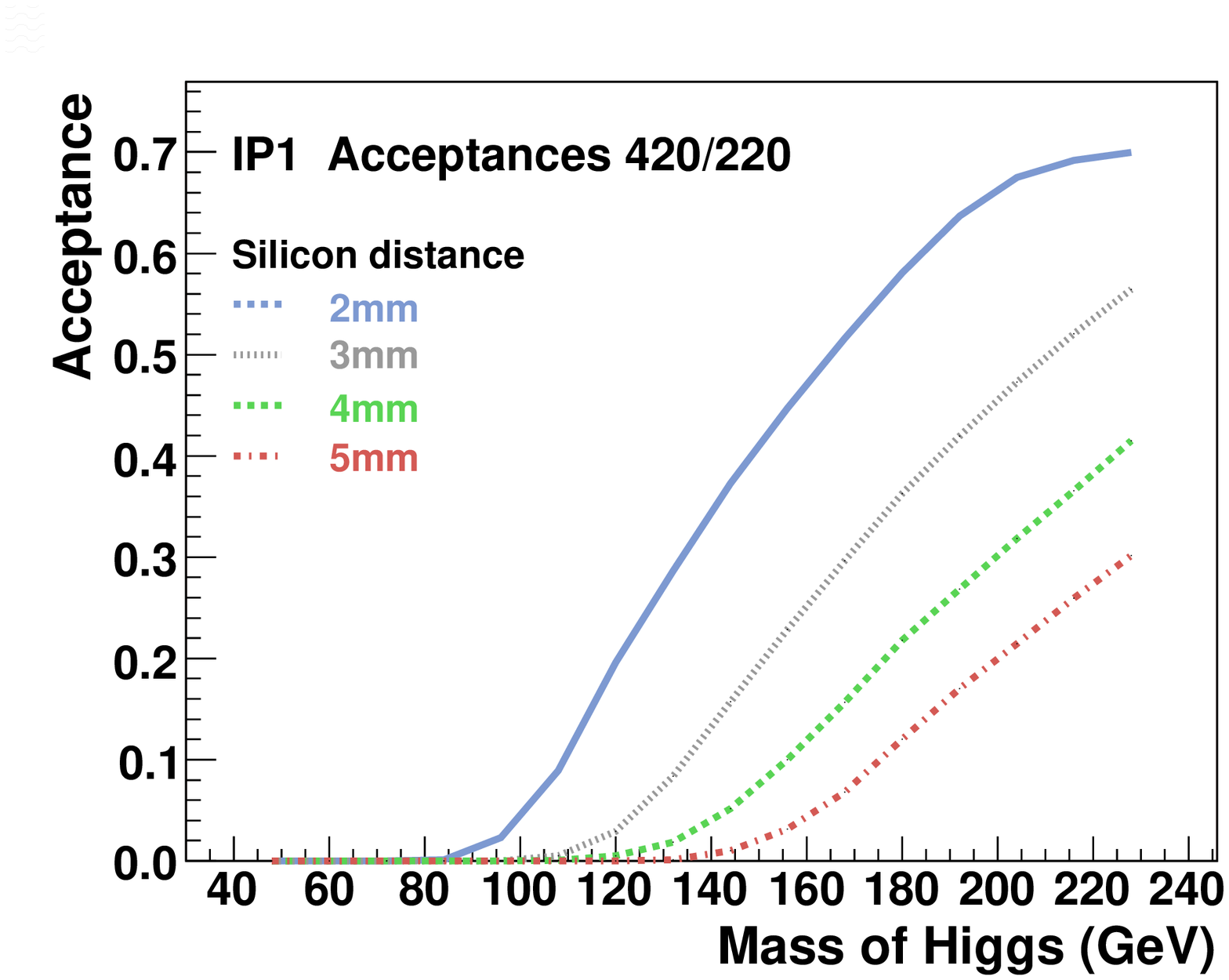}
        \caption{Acceptance of forward tagging systems as function of 
mass of the centrally produced object, taken here as a Higgs.}
        \label{fig:fwd_8}
\end{figure}

To perform the tracking of the protons into the relevant detector
regions, two programs (FPtrack and Hector) have been written for ATLAS
and CMS respectively \cite{Track}. They enable us to evaluate the
acceptance of the apparatus under various conditions, as 
illustrated in Fig.~\ref{fig:fwd_8}.  The 420 m systems used on their
own provide substantial acceptance for exclusively produced masses up
to approximately 150 GeV/c$^2$, and even if the silicon can be moved
only to 7 mm from the beam, the acceptance at the critical region of
120 GeV/c$^2$ is not affected. By using the 420 m systems in
conjunction with those at 220 m, a greatly extended mass range is
achieved with excellent acceptances.

The mass $M_X$ of an exclusively
produced final state can be evaluated by reconstructing the momenta
of the forward protons; this is achievable by means of
polynomial-based formulae in terms of the horizontal position and
angle in the detector regions.  The value of $M_X$ is then $2\sqrt{(p_0
- p_1)(p_0-p_2)}$ for an incoming beam momentum $p_0$ and outgoing
proton momenta $p_1,\,p_2$.  Various uncertainties smear out this
calculation, notably the intrinsic spread on $p_0$. 
A mass uncertainty of 3-4 GeV/$c^2$ is obtainable.  
This is nearly always 
better than the uncertainty obtained by direct measurement in the
central detector.  An exception is when the central state
consists of two photoproduced muons.  This promises to be a key
process which we intend to use to calibrate the proton momentum
measurements.

Pile-up backgrounds are a potential problem if there are many
interaction verteces in a single beam crossing.  To identify the
correct vertex, very precise timing devices, based on Cherenkov
radiation detection, will be installed in the forward detection regions.  These
are currently under study.

\section{Conclusions}
For more details, the  full FP420 project report may be consulted \cite{Albrow}.
Forward tagging opens up a wide range of diffraction and
photoproduction processes at LHC.  There is discovery potential in
some cases, while in others, known processes can be studied in more
depth. This is a major new area of physics potential for the LHC.


\end{document}